\documentclass{iaus}
\usepackage{graphicx}

\usepackage{natbib}
\title{On the origin of complex stellar populations in star clusters}
\author[Jan Pflamm-Altenburg \& Pavel Kroupa]
{Jan Pflamm-Altenburg$^{1,2}$ and Pavel Kroupa$^{1,2}$}

\affiliation{$^1$Argelander Institut f\"ur Astronomie (AIfA),
Auf dem H\"ugel 71, D-53121 Bonn, Germany
\break email: jpflamm@astro.uni-bonn.de, pavel@astro.uni-bonn.de\\[\affilskip]
$^2$ Rhine Stellar Dynamics Network (RSDN)}

\pubyear{2007}
\volume{246}  
\pagerange{}
\date{28. August 2007 and in revised form ??}
\jname{Dynamical Evolution of Dense Stellar Systems}
\editors{E. Vesperini (Chief Editor), M. Giersz, A. Sills}

\begin{document}
\maketitle
\begin{abstract}
The existence of complex stellar populations in some star clusters 
challenges the understanding of star formation. E.g.
the ONC or the sigma Orionis cluster host much older stars 
than the main bulk of the young stars. 
Massive star clusters ($\omega$ Cen, G1, M54) 
show metallicity spreads corresponding to 
different stellar populations with large age gaps.
We show that (i) during star cluster formation 
field stars can be captured and (ii) very massive globular clusters
can accrete gas from a long-term embedding inter stellar medium 
and restart star formation.
\end{abstract}
\section{Field star capture}
The collapsing cloud is described by a Plummer potential
with a fixed total gas mass and a time-dependent 
Plummer parameter, infinite at the beginning 
and decreasing within the collapse time-scale. 
The field stars are simulated by test particles
which are initially uniformly distributed in a sphere 
and have a Gaussian velocity distribution. 
\subsection{ONC}
\begin{figure}
  \includegraphics[width=0.5\textwidth]{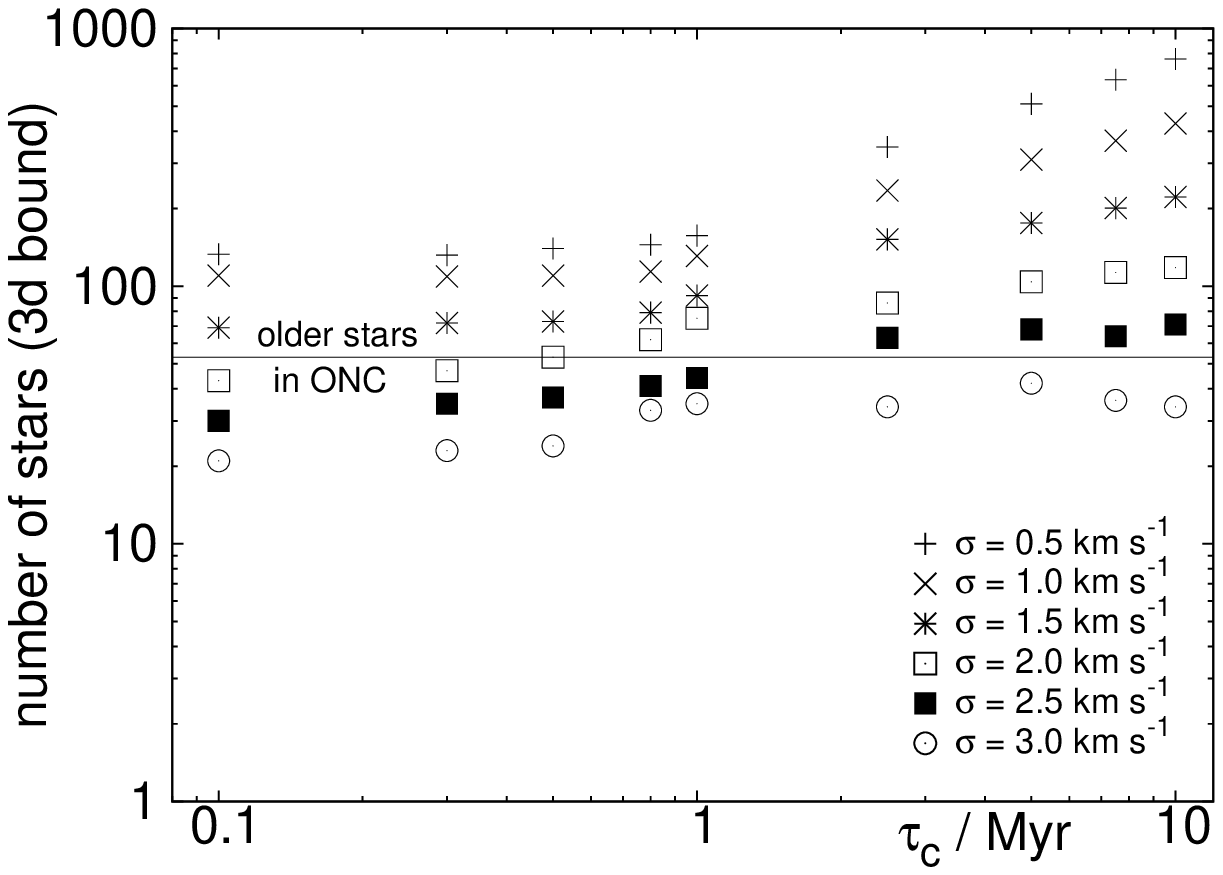}
  \includegraphics[width=0.5\textwidth]{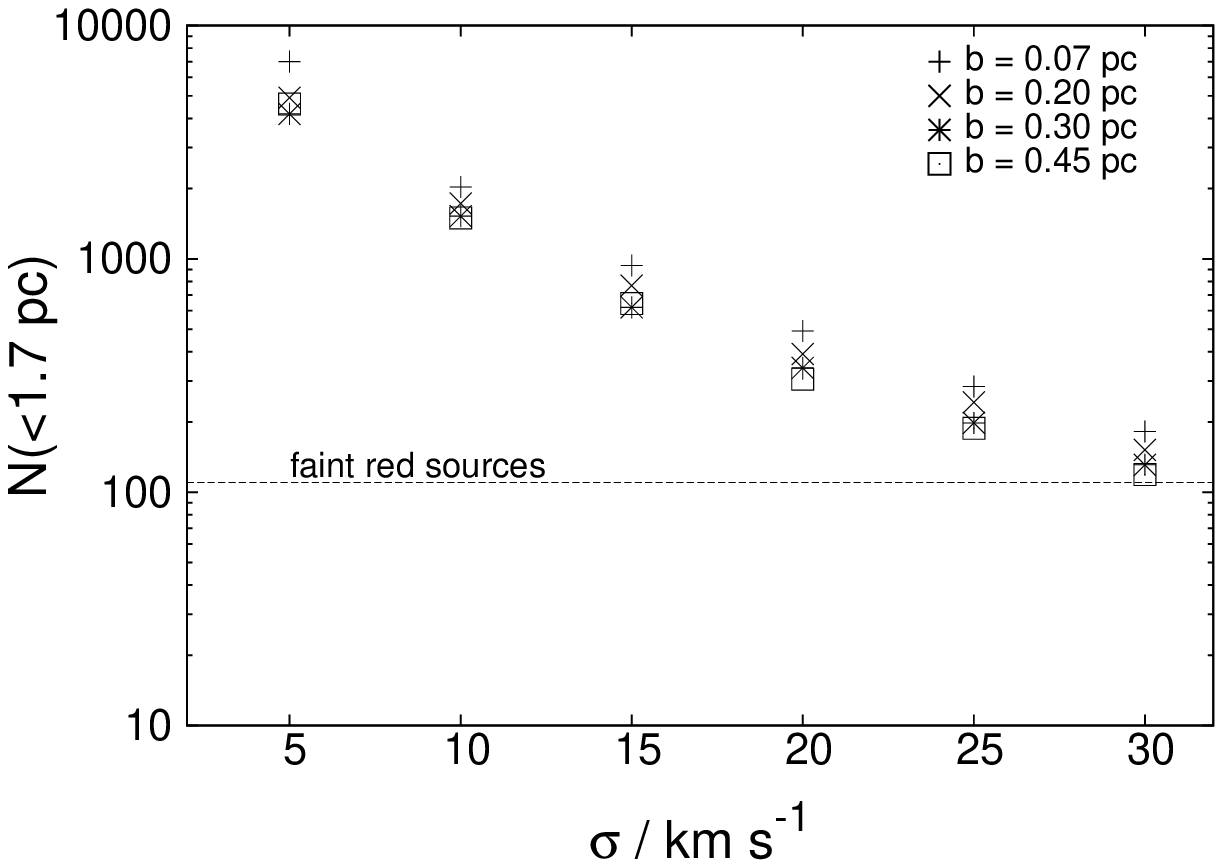}
  \caption{{\it Left}: Captured stars in the 
    ONC within a radius of 2.5~pc for different velocity dispersions, $\sigma$, 
    and collapse time-scales, $\tau_\mathrm{c}$,  \citep{pflamm-altenburg2007a}. 
    {\it Right}: Field stars in the central sphere of 1.7~pc radius in 
    R136 for different velocity dispersions and final 
    Plummer parameters, $b$,
    \citep{pflamm-altenburg2007b}.}
    \label{fig01_02}
\end{figure}

\citet{palla2005a} found 4 stars out of 84 low
mass stars in the ONC being 10~Myr older than the main
bulk of stars and followed that star formation is prolonged. Extrapolating, 
$\approx$~53 such older stars are expected in the whole ONC
\citep{pflamm-altenburg2007a}. The number of simulated 
captured stars within a radius of 2.5~pc of the centre of the ONC
is plotted in Fig.~\ref{fig01_02} (left) for different collapse time-scales
and initial background velocity dispersions. Age spreads in young star 
clusters may therefore
not be due to prolonged star formation but can be explained by stellar 
capture.
\subsection{R136}
\citet{brandl1996a} found 110 faint red sources in a field of 
3$\times$3~pc$^2$ of the centre of R136 in the LMC of unknown physical nature
and excluded that they can be red giants as their required age would be 
larger than 350~Myr. 
As membership probabilities can not be determined for stars in R136 all stars within
the central sphere with a radius of 1.7~pc 
are calculated after the collapse with a time-scale
of 10~Myr has stopped \citep{pflamm-altenburg2007b}. 
The faint red sources can indeed be captured old red giants.

\section{Gas accretion}  
\begin{figure}
  \begin{center}
    \includegraphics[width=0.5\textwidth]{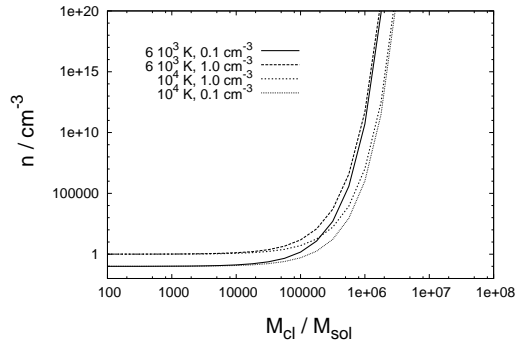}
    \end{center}
  \caption{Particle density of the ISM at the origin of 
    the cluster potential as a function of the total cluster 
    mass \citep{pflamm-altenburg2007c}. Note the instability 
    near 10$^6$~M$_\odot$.}
    \label{fig03}
\end{figure}

Globular clusters more massive than $\approx$~10$^6$~M$_\odot$
such as $\omega$~Cen, G1 and M54 show a spread in metallicity
corresponding to different stellar population  with age gaps of several hundred
Myr or a few Gyr. M54 and $\omega$~Cen are confirmed and supposed, 
respectively, to have been embedded in the ISM of a dwarf galaxy.
To explore the effect of the cluster potential on a long-term embedding
co-moving ISM we calculate the hydrostatic solution of an isothermal 
non-self-gravitating gas with an additional Plummer potential
\citep{pflamm-altenburg2007c}.
The central particle density in the cluster potential is plotted in   
Fig.~\ref{fig03} for different temperatures and densities of the warm component
of the ambient ISM. Star clusters more massive than $\approx$~10$^6$~M$_\odot$
may cause an instability of the ISM, start gas accretion and form stars.
Gas accretion can explain multiple stellar 
populations in massive star clusters and the change of the mass-radius
relation at cluster masses of $\approx$~10$^6$~M$_\odot$.

\begin{thebibliography}{}

\bibitem[\protect\citeauthoryear{{Brandl}, {Sams}, {Bertoldi}, {Eckart},
  {Genzel}, {Drapatz}, {Hofmann}, {Loewe} \& {Quirrenbach}}{{Brandl}
  et~al.}{1996}]{brandl1996a}
{Brandl} B.,  {Sams} B.~J.,  {Bertoldi} F.,  {Eckart} A.,  {Genzel} R.,
  {Drapatz} S.,  {Hofmann} R.,  {Loewe} M.,    {Quirrenbach} A.,  1996, 
  \textit{ApJ},
  466, 254

\bibitem[\protect\citeauthoryear{{Palla}, {Randich}, {Flaccomio} \&
  {Pallavicini}}{{Palla} et~al.}{2005}]{palla2005a}
{Palla} F.,  {Randich} S.,  {Flaccomio} E.,    {Pallavicini} R.,  2005, 
\textit{ApJ} (Letters),
  626, L49

\bibitem[\protect\citeauthoryear{{Pflamm-Altenburg} \&
  {Kroupa}}{{Pflamm-Altenburg} \& {Kroupa}}{2007a}]{pflamm-altenburg2007b}
{Pflamm-Altenburg} J.,  {Kroupa} P.,  2007a, in preparation

\bibitem[\protect\citeauthoryear{{Pflamm-Altenburg} \&
  {Kroupa}}{{Pflamm-Altenburg} \& {Kroupa}}{2007b}]{pflamm-altenburg2007a}
{Pflamm-Altenburg} J.,  {Kroupa} P.,  2007b, \textit{MNRAS}, 375, 855

\bibitem[\protect\citeauthoryear{{Pflamm-Altenburg} \&
  {Kroupa}}{{Pflamm-Altenburg} \& {Kroupa}}{2007c}]{pflamm-altenburg2007c}
{Pflamm-Altenburg} J.,  {Kroupa} P.,  2007c, in preparation

\end{thebibliography}

\end{document}